# Multi-Time-Scale Input Approaches for Hourly-Scale Rainfall-Runoff Modeling based on Recurrent Neural Networks

Kei Ishida[1,2], Masato Kiyama[3], Ali Ercan[4], Motoki Amagasaki[3], Tongbi Tu[5]

**Abstract**

This study proposes two straightforward yet effective approaches to reduce the required computational time of the training process for time-series modeling through a recurrent neural network (RNN) using multi-time-scale time-series data as input. One approach provides coarse and fine temporal resolutions of the input time-series to RNN in parallel. The other concatenates the coarse and fine temporal resolutions of the input time-series data over time before considering them as the input to RNN. In both approaches, first, finer temporal resolution data are utilized to learn the fine temporal scale behavior of the target data. Next, coarser temporal resolution data are expected to capture long-duration dependencies between the input and target variables. The proposed approaches were implemented for hourly rainfall-runoff modeling at a snow-dominated watershed by employing a long and short-term memory (LSTM) network, which is a newer type of RNN. Subsequently, the daily and hourly meteorological data were utilized as the input, and hourly flow discharge was considered as the target data. The results confirm that both of the proposed approaches can reduce the computational time for the training of RNN significantly (up to 32.4 times). Furthermore, one of the proposed approaches improves the estimation accuracy.

## 1. Introduction

Recurrent neural networks (RNNs) are gathering attention for time-series modeling in various domains, including hydrology. RNN is a variant of a deep neural network with a specific architecture. Specifically, it can receive sequential data one by one as the input, and then generate outputs by considering the sequence of the input data. Owing to this structure, RNN can learn the dependencies between the input and targeted data sequences. However, the traditional RNN has the limitation of learning long-term dependencies known as the vanishing gradient problem. To address the vanishing gradient problem, a new variant of RNN was developed by incorporating components such as the cell state and input, output, and forget gates, which is called the long short-term memory (LSTM) network (Gers et al., 2000; Hochreiter and Schmidhuber, 1997). These components enable LSTM to learn long-term dependencies between the input and targeted data sequences. Due to this capability of LSTM, RNN has gained attention for time-series modeling in hydrology.

In the last decade, several studies have utilized LSTM, including its sequence-to-sequence extension, for time-series modeling and forecasting in hydrology. LSTM has frequently been applied to river flow discharge modeling (Kratzert et al., 2019) and forecasting (Kao et al., 2020; Li et al., 2020; Liu et al., 2020; Song et al., 2019; Tian et al., 2018; Xiang et al., 2020; Zhu et al., 2020). In addition, groundwater modeling and forecasting are prominent topics in time-series modeling using LSTM (Bowes et al., 2019; Jeong et al., 2020; Jeong and Park, 2019; Zhang et al., 2018). Some research groups have utilized LSTM for statistical downscaling of daily precipitation (Miao et al., 2019; Misra et al., 2018; Tran Anh et al., 2019). Also, LSTM has been employed for other time series modeling in hydrology, such as lake water level forecasting (Hrnjica and Bonacci, 2019; Liang et al., 2018), reservoir operation modeling (Zhang et al., 2018), and tsunami flood forecasting (Hu et al., 2019). These studies exhibit a substantial scope of RNN (LSTM) to model time-series issues in hydrology.

1. CWMD, Kumamoto University
2. IROAST, Kumamoto University
3. FAST, Kumamoto University
4. Department of Civil and Environmental Engineering, University of California, Davis
5. School of Civil Engineering, Sun Yat-Sen University

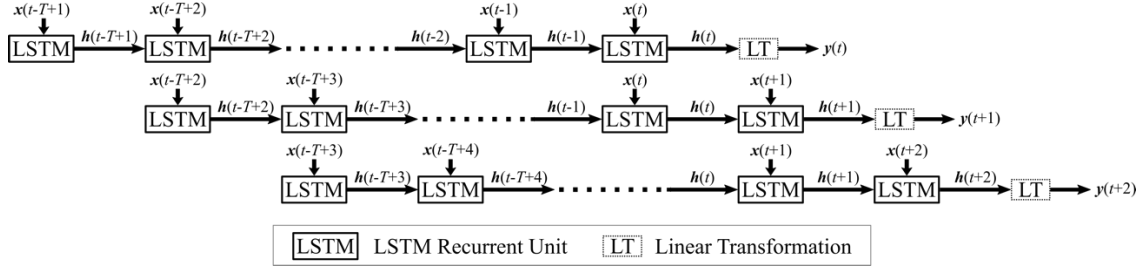

Figure 1. Temporal relationship between the input and output time-series of LSTM.

In contrast, RNN, including LSTM, has a limitation while modeling time-series. The computational requirements of RNNs are strongly influenced by the length of the input data sequence (IDL). Notably, IDL for RNN is different from the length of time (duration) for time-series modeling. RNN deals with time-series data as sequential data without considering time, although it can consider the order of the input data sequences. This implies that IDL becomes much longer with a finer temporal scale of time series data as the input. For example, Kratzert et al. (2018) used 365 days of meteorological variables as the input to LSTM to model daily rainfall-runoff in snow-dominated watersheds to reflect long-term dependencies between meteorological time-series variables and flow discharge. When modeling rainfall-runoff on an hourly scale, IDL becomes 8760 (365 days × 24 h). Meanwhile, RNN requires tuning various model options known as hyper-parameters in machine learning, which involves many trial-and-errors to optimize the hyper-parameters. In addition, using RNN to model time-series at a finer temporal scale would increase computational resources unreasonably when considering a long-duration of dependencies between the time-series of input and target variables.

To this end, this study proposes two approaches to reduce IDL for time-series modeling by RNN when there exist long-duration dependencies between the input and target variables. To generate outputs that exhibit behavior at a certain temporal scale, a time-series model requires the input time-series data following the same temporal resolution. However, it appears unnecessary to use the same temporal resolution of the input time-series data to reflect the long-duration dependencies on the outputs. In this regard, a coarser resolution of the input time-series data can be sufficient. Therefore, this study proposes two straightforward yet effective approaches for RNNs that use multi-temporal-scale input time-series data. Specifically, both proposed approaches utilize the same time-series variables at a finer time resolution and a coarser time resolution as the input.

The use of input time-series data at a coarser time resolution is anticipated to facilitate the model to learn dependencies between the input and target time-series variables for a long-duration. In contrast, these at a finer time resolution are anticipated to catch a fine temporal scale behavior of the target data for a short time duration.

In this study, LSTM among RNNs was employed because LSTM is expected to have advantages in capturing long-term dependencies. Both proposed approaches are implemented for hourly rainfall-runoff modeling at a snow-dominated Ishikari River watershed (IRW). In addition, the accuracy and computational resource requirements of both proposed approaches are compared with each other and with the classical approach. For both approaches, the daily and hourly time-series of meteorological variables were utilized as inputs. The general approach refers to LSTM with the hourly temporal resolution of the input variables.

## 2. Methodology

LSTM is used to model time-series. Since LSTM has several structural variants, LSTM used in this study has a cell state $c$ and three gates: the cell gate $g_c$, input gate $g_i$, output gate $g_o$, and forget gate $g_f$,

$$g_i(s) = \sigma(W_{ii}x(s) + b_{ii} + W_{hi}h(s-1) + b_{hi}), \quad (1)$$

$$g_f(s) = \sigma(W_{if}x(s) + b_{if} + W_{hf}h(s-1) + b_{hf}), \quad (2)$$

$$g_o(s) = \sigma(W_{io}x(s) + b_{io} + W_{ho}h(s-1) + b_{ho}), \quad (3)$$

$$c(s) = g_f(s) \otimes c(s-1) + g_i(s)$$
$$\otimes \tanh(W_{ic}x(s) + b_{ic} + W_{hc}h(s-1) + b_{hc}), \quad (4)$$

$$h(s) = g_o(s) \otimes \tanh(c(s)), \quad (5)$$

where $x(s)$ and $h(s)$ denote the input vector and hidden state at time $s$, respectively. $W$ and $b$ are the weights and biases, respectively. Notably, two subscripts of each weight and bias indicate the

a) ParaLSTM

b) ConcLSTM

Figure 2. Input time-series. (a) for ParaLSTM and b) for ConcLSTM.

input/hidden state vector and gate, respectively. The weights and biases are the learnable parameters that need to be tuned. Also, σ and ⊗ represent the sigmoid function and Hadamard product, respectively.

When the length and end time of the input vector time-series are set to $T$ and $t$, respectively, Equations 1–5 are recurrently used from $s = t − T + 1$ to $t$ with the input time-series vector $x(s)$, resulting in the hidden state $h(t)$ at time $t$. Subsequently, the hidden state $h(t)$ is linearly transformed and assigned to an activation function when the activation function produces the output vector $y(\varsigma)$ at time $\varsigma$. Here, for time-series modeling, $\varsigma$ was set equal to $t$ (Figure 1), and the linear function was utilized as the activation function. To obtain a time-series of the output vector during the study period, the aforementioned procedure was repeated from the beginning to the end of the study period. Notably, LSTM can learn long-term dependencies between the input and output time-series; however, the duration of dependencies learned by LSTM depends on IDL ($T$).

Therefore, firstly, for time-series modeling, the temporal increment of the input time-series is considered the input to LSTM, which is generally the same as that of the time-series of the output vector. Secondly, for hourly time-series modeling, the input time-series at the hourly scale ($x = x^H$) was used. Notably, while setting IDL ($T$) to $T^H$, time-series of the hourly scale input variable ($x^H(s): s = t^H − T^H + 1, t^H − T^H + 2, \cdots, t^H − 1, t^H$) is used as the input to LSTM. Subsequently, LSTM is executed hourly to generate an hourly output vector $y(t^H)$. In this study, this original approach of hourly time-series modeling is referred to as OrigLSTM.

The first proposed approach provides hourly and daily time-series parallel to LSTM as the input, which is referred to as ParaLSTM. When the length of the hourly and daily input time-series (HIDL and DIDL) are set to $T^H$ and $T^D$, respectively, the hourly $x^H$ and the daily input time-series $x^D$ are defined as:

$$x^H(s^H) = \begin{bmatrix} x_1^H(s^H) \\ x_2^H(s^H) \\ \vdots \\ x_n^H(s^H) \end{bmatrix}$$

$(s^H = t^H − T^H + 1, t^H − T^H + 2, \cdots, t^H − 1, t^H),$

$$x^D(s^D) = \begin{bmatrix} x_1^D(s^D) \\ x_2^D(s^D) \\ \vdots \\ x_n^D(s^D) \end{bmatrix}$$

$(s^D = t^D − T^D + 1, t^D − T^D + 2, \cdots, t^D − 1, t^D),$

where $n$ is the number of types of input variables. $s^H$ and $s^D$ denote the times for the hourly and daily input time-series, respectively. The units of $s^H$ and $s^D$ are hour and day, respectively. $t^H$ and $t^D$ represent the end times of the input time-series at the hourly and daily scales, respectively. The time of the outputs was $t^H$, and $t^D$ was the date of $t^H$.

When $T^H = T^D$, the hourly and daily input time-series are arranged in parallel and provided as the input to LSTM, as shown in Figure 2a. In contrast, when $T^H \neq T^D$, a special treatment is required to use these time series as the input to LSTM, which can receive a single IDL. In natural language processing, the padding method is frequently used to align IDLs for RNN. Similarly, Ishida et al. (2020) incorporated a yearly time-series of global air temperature together with hourly input time-series for LSTM by employing this method to reflect the effects of global warming on the hourly scale coastal sea level modeling. In addition, this study utilizes the padding method to provide hourly and daily time-series together as an input to LSTM. This study considered only the case of $T^D > T^H$. Then, IDL ($T$) to ParaLSTM is equal to $T^D$. Thus, the hourly and daily time-series are provided as the input to LSTM parallelly. The gap between $T^H$ and $T^D$ is padded by a specific value to align their lengths. Generally, zero is used for adding after normalizing the input data. The time-series of the input vectors is shown in Figure 2a. This approach exhibits a temporal inconsistency between the hourly and daily input time-series.

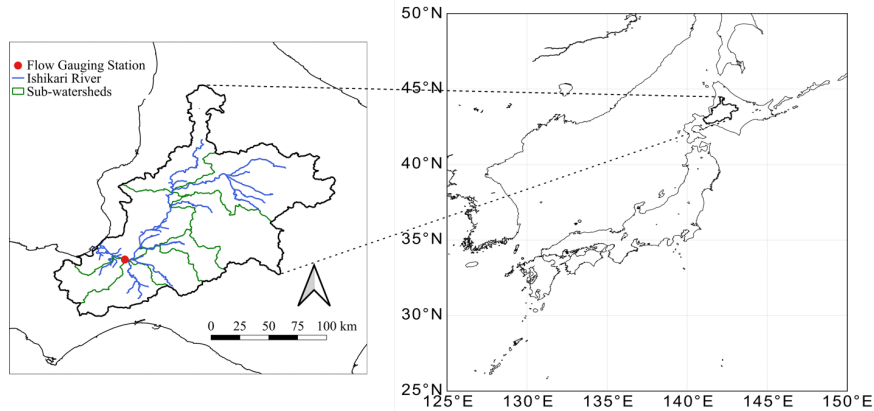

Figure 3. Study area

The second approach, referred to as ConcLSTM, concatenates the hourly and daily input vectors along the time axis, and uses these time-series together. However, before concatenating the time-series, the overlapping period between the hourly and daily time-series was removed. The integer part of $T^H/24$ was set to $T^{D0}$. ConcLSTM uses a part of the daily input time-series between $t^D - T^D + 1$ and $t^D - T^{D0}$, as shown in Figure 2b.

Both approaches can significantly reduce the length of the input time-series. For instance, ParaLSTM requires merely 365 length input time-series to incorporate one-year information of the input time-series into a model. It is 1/24 of the length required by OrigLSTM, although the number of input variables increased twofold. ConcLSTM requires a longer input time-series than ParaLSTM, which is still much shorter than OrigLSTM. Notably, both approaches are expected to significantly reduce computation time because of the shorter length of the input time-series. As shown in Figure 2a, ParaLSTM has an inconsistency in time between the hourly and daily input time-series. This approach is utilized based on the justification that LSTM can learn relationships between the input time-series and target data even when temporal inconsistency exists. In contrast, there is almost no temporal inconsistency between these time-series in ConcLSTM. Although its temporal increment suddenly changes from an hourly to a daily scale, the input data sequence is arranged in the temporal order.

3. **Case Study**

3.1. **Hourly-scale Rainfall-runoff Modeling**

As a case study, we selected hourly rainfall-runoff modeling at a snow-dominated watershed to investigate the potential of the aforementioned approaches for time-series modeling. Since it can potentially require a long duration of the input data to reflect the effects of snow accumulation and melting to flow discharge by LSTM, hourly scale rainfall-runoff modeling at a snow-dominated watershed is appropriate for the analysis. Then, rainfall-runoff models were developed based on the three approaches with daily and hourly input time-series. In addition, LSTM with only an hourly input time-series is utilized to create a rainfall-runoff model. Finally, the rainfall-runoff models with three approaches were compared to investigate their potential.

3.2. **Study Area and Data**

As a study watershed, we targeted hourly flow discharge at IRW in the Hokkaido region at the northern end of Japan. The Ishikari River is the third-longest river in Japan, with a length of 268 km, originating from the mountain peak of 1,967 m and flowing into the sea. The catchment area of IRW is 14,330 km² that is the second largest in Japan. Geographically, IRW is located in a cold region and receives snowfall from October to March, especially in high-elevation areas. This snowpack melts during the spring to early summer, i.e., from March to June.

Ishikari Ohashi gauging station was selected to obtain the hourly flow discharge for the target data of the model. As shown in Figure 3, this station is 26.60 km from the outlet along the Ishikari River. The hourly flow discharge data were obtained from the Water Information System (WIS) driven by the Ministry of Land, Infrastructure, Transport, and Tourism of Japan (http://www1.river.go.jp/). Notably, data were available from 1998 to 2016, with some missing parts in the hourly flow discharge data. These missing parts were removed from the target data while training and from data while calculating the evaluation metrics.

This study utilized hourly precipitation and air temperature as inputs. Because the catchment area of IRW is relatively large, the spatial variabilities of precipitation may not be negligible within IRW. Therefore, we use the spatial average of the hourly precipitation data at the sub-regions of IRW as the input. These sub-regions are illustrated in Figure 3. Hourly precipitation data were obtained from the radar raingauge-analyzed precipitation (RRAP) generated by the Japan Meteorological Agency. RRAP, available from 1988 to the present, is a gridded precipitation data obtained by modifying the radar precipitation with gauging precipitation data. However, the provided grid resolution has gradually improved throughout. Currently, RRAP is provided at a 1-km spatial resolution from 2006 to the present. This study utilized RRAP at a 1-km spatial resolution. To generate the input time-series data, the spatially averaged values of RRAP were calculated for the sub-regions of IRW.

Many meteorological observation stations are located within IRW, which are operated by the Japan Meteorological Agency. However, this study utilized hourly air temperature data only at a single station due to data availability. At some stations, air temperature data were available either for recent or older period. In addition, many missing data points were present at most stations. Air temperature data are often highly correlated among nearby stations. Thus, the input data are normalized for neural networks, including LSTM. When data are highly correlated among the stations within a watershed, the air temperature data at a single station can be considered representative for time-series modeling with a neural network, unlike physical or conceptual models. Thus, a station was selected among the stations within the watershed where the missing parts were minimum. At the selected station, merely seven hours were missing during 11 years from 2006 to 2016. These missing parts were imputed by linear interpolation.

### 3.3. Model Implementation

LSTM has several tuning options called hyperparameters. Notably, for comparisons, the same configuration was used in the three approaches. The hidden state length was set at 50. Only a single layer of the LSTM recurrent unit was utilized. HIDL was set to 8760 (hours) for OrigLSTM, and DIDL was set to 365 (days) for both ParaLSTM and ConcLSTM. Then, several HIDLs were tested for ParaLSTM and ConcLSTM: 24, 48, and 120 (hours). The given dataset was segregated into three sets: training, validation, and test datasets. Then, the model was calibrated using the training and validation datasets, and it was verified with the test dataset as detailed below.

The model parameters: weights and biases, are updated using the training dataset by means of the back-propagation approach. An update iteration is performed with a subset of the training dataset known as a batch or mini-batch. Notably, these batches are generated using the shuffle sampling method. Specifically, this method randomly extracts samples from the training dataset for each batch with no duplication until all samples are selected. When all samples were selected, the aggregation of the update iterations was considered an epoch. The sample size of each batch or batch size was set to 256. The back-propagation approach is implemented using the gradient descent method with an optimization algorithm. The mean square error (MSE) was selected as the loss function for the gradient descent method. The error calculated by this function is called loss. Adaptive moment estimation (Adam; Kingma and Ba, 2014) is employed as the optimization algorithm to adjust the gradient during each update iteration.

The validation dataset is utilized with an early stopping criterion to avoid the overfitting of the model, and save computation resources for model calibration. At each epoch, the network whose parameters were updated when the training dataset was executed with the validation dataset, and then the loss was calculated for the validation dataset. When this loss continuously increases during a specific number of epochs, it is considered that the model overfits the training dataset. Then, the training stops and the parameters that yielded the minimum loss for the validation dataset were selected. The number of epochs to determine whether the overfitting occurs is called patience, which was set to 50 in this study.

Table 1 Statistical comparison among the three approaches. NEpochs and Time/EP for the training, and RMSE, R, NSE of the final trained results obtained by OrigLSTM, ParaLSTM, and ConcLSTM with several HIDLs.

|  | HIDL | NEpochs | Time/EP (s) | Training | | | Validation | | | Test | | |
|---|---|---|---|---|---|---|---|---|---|---|---|---|
|  |  |  |  | RMSE (m³/s) | R | NSE | RMSE (m³/s) | R | NSE | RMSE (m³/s) | R | NSE |
| OrigLSTM | 8760 | 61.4 ± 49.3 | 58.7 ± 0.17 | 170.2 | 0.930 | 0.859 | 202.2 | 0.896 | 0.789 | 222.4 | 0.884 | 0.771 |
| ParaLSTM | 24 | 48.2 ± 22.5 | 2.7 ± 0.01 | 135.7 | 0.955 | 0.911 | 197.4 | 0.897 | 0.799 | 233.0 | 0.867 | 0.749 |
|  | 48 | 43.7 ± 24.4 | 2.7 ± 0.01 | 141.3 | 0.956 | 0.903 | 190.0 | 0.909 | 0.813 | 214.1 | 0.889 | 0.788 |
|  | 120 | 41.2 ± 26.1 | 2.7 ± 0.01 | 137.3 | 0.954 | 0.909 | 172.8 | 0.923 | 0.846 | 186.8 | 0.918 | 0.838 |
| ConcLSTM | 24 | 59.1 ± 38.3 | 2.7 ± 0.01 | 134.2 | 0.957 | 0.913 | 230.7 | 0.857 | 0.725 | 243.5 | 0.855 | 0.725 |
|  | 48 | 37.1 ± 21.3 | 2.8 ± 0.01 | 156.2 | 0.943 | 0.882 | 202.2 | 0.889 | 0.789 | 244.6 | 0.867 | 0.723 |
|  | 120 | 82.4 ± 42.8 | 3.2 ± 0.01 | 123.8 | 0.964 | 0.926 | 192.1 | 0.902 | 0.809 | 233.5 | 0.872 | 0.748 |

Generally, weights and bias of LSTM are initialized using random values because randomness in the initial states of parameters can affect the model accuracy. Therefore, the aforementioned calibration process was performed 100 times for each approach with each HIDL. The best-trained model for each LSTM with each HIDL was extracted with respect to the losses for the validation period. Finally, the best-trained model was verified with the test dataset. All computations were conducted using a computer with 64 RAM, Intel Core i7-10700k, and NVIDIA GeForce RTX 2080 Ti. Models were implemented using Python and its deep learning framework Pytorch (Paszke et al., 2019).

Both proposed approaches, ParaLSTM and ConcLSTM, were compared with OrigLSTM based on the model accuracy and required training time. The model accuracy was investigated using three evaluation metrics: root mean square error (RMSE), correlation coefficient (R), and Nash–Sutcliffe efficiency (NSE). There are 100 trained results for each approach with each HIDL for the training and validation periods. Boxplots were created to compare the evaluation metrics for the training and validation periods. In contrast, there was a single value for the test period. Each metric was tabulated for comparison. In addition, the required time for training was compared using two measures. First, required computational time for training at each epoch (Time/EP), and the second, the number of epochs taken by each training process to reach the minimum loss for the validation period for each training process (NEpochs).

## 4. Results

Time/EP and NEpochs for the training process were obtained to examine the computational efficiency of the proposed approaches, as shown in Table 1. The standard deviation of the computational time was less than 1% of the average computational time for each configuration. The difference in the computational time among epochs is negligible. In ParaLSTM, IDL ($T$) is equal to DIDL ($T^H$) when HIDL ($T^H$) ≤ DIDL ($T^D$) because DIDL ($T^D$) is set to 365 for all cases. Consequently, Time/EP is 2.7 s for an epoch on average. In ConcLSTM, IDL ($T$) gradually increased with HIDL ($T^H$). In addition, Time/EP increases with HIDL ($T^H$), although the rate of increase with HIDL is not linear. While Time/EP for HIDL ($T^H$) = 24 is 2.7 s on average, it is 3.2 s for HIDL = 120. Time/EP for OrigLSTM is 58.7 s on average. ParaLSTM is approximately 21.7 times faster than OrigLSTM when considering the same time duration (365 days = 8769 h) for the input. Even ConcLSTM with HIDL = 120, which is the slowest of the proposed method with all the HIDLs, is approximately 18.3 times faster than OrigLSTM with respect to Time/EP.

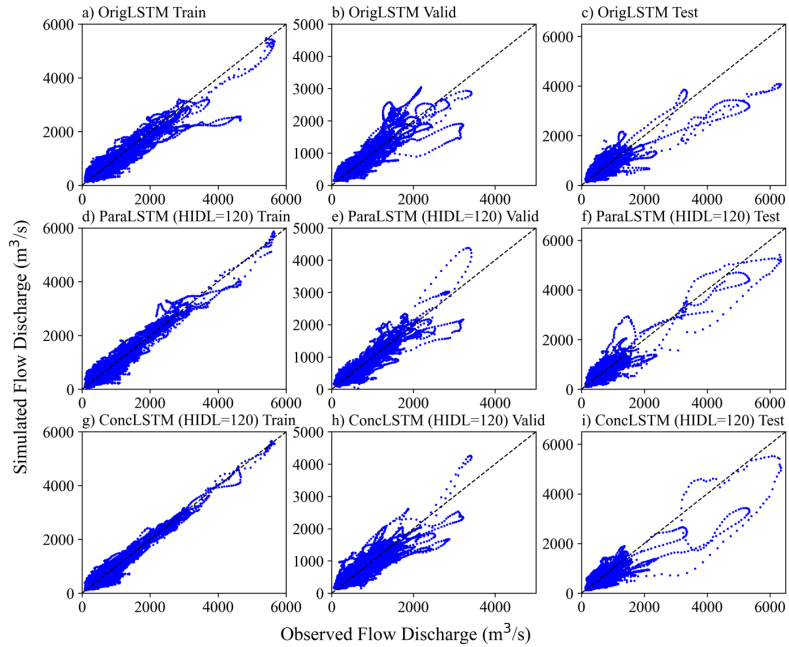

Figure 4. Scatter plots between the hourly observed and simulated flow discharges of the final trained results obtained by OrigLSTM, ParaLSTM with HIDL = 120, and ConcLSTM with HIDL = 120.

The average of NEpochs is relatively smaller for ParaLSTM than for the others. The average of NEpochs is 61.4 for OrigLSTM, whereas, for ParaLSTM, the average of NEpochs gradually decreases from 48.2 to 41.2, with a longer HIDL. There is no clear relationship between the average of NEpochs and HIDL for ConcLSTM. It is the smallest (37.1) with HIDL = 48, and the largest (82.4) with HIDL = 120 for ConcLSTM. In addition, the standard deviation of the NEpochs is between 22.5 and 26.1 for ParaLSTM, which is relatively small. In contrast, the standard deviation is 49.3 for OrigLSTM. It fluctuates with ConcLSTM between 21.3 and 42.8. Also, except for HIDL = 48, the standard deviation of ConcLSTM is larger than that of ParaLSTM.

The model accuracy was compared with the three approaches using three evaluation metrics: RMSE, R, and NSE. To reiterate, the final trained result was obtained from the best results for each configuration based on the loss for the validation period. These evaluation metrics of the final trained result for each configuration are listed in Table 1. ParaLSTM, except for HIDL = 24, yielded better model accuracy for the test period compared to OrigLSTM, whereas ConcLSTM with all the HIDL yielded less model accuracy for the test period. ParaLSTM and ConcLSTM yield the best model accuracy with HIDL= 120 based on these evaluation metrics. One-to-one plots between the observed and simulated flow discharges obtained by OrigLSTM, ParaLSTM with HIDL=120, and ConcLSTM with HIDL=120 are delineated in Figure 4. ParaLSTM generated better estimations of the large flow discharge during the test period (Figure 4f) compared to OrigLSTM (Figure 4c) and ConcLSTM (Figure 4i). As shown in Table 1, ParaLSTM with HIDL=120 improved RMSE, R, and NSE by 29.8 $m^3$/s, 0.029, and 0.057, respectively, for the test period, compared to OrigLSTM. In contrast, these evaluation metrics obtained by ConcLSTM with HIDL=120 are 11.1 $m^3$/s, 0.012, and 0.023, respectively, which are worse than those obtained by OrigLSTM.

These evaluation metrics obtained by ParaLSTM with all the HIDLs were better than OrigLSTM for the training and validation periods. Although ConcLSTM yields better model accuracy merely with HIDL = 120 for the validation period compared to OrigLSTM. These evaluation metrics by ConcLSTM with all the HIDLs were better than OrigLSTM for the training period. Notably, these values were comparable to those of ParaLSTM. For example, ParaLSTM with HIDL=120 improved RMSE, R, and NSE values by 32.9 $m^3$/s, 0.024, and 0.05, respectively, for the training period, and 29.4 $m^3$/s, 0.027, and 0.057, respectively, for the validation period. Meanwhile, the differences in RMSE, R, and NSE between OrigaLSTM and ConcLSTM with HIDL = 120 were 46.4 m$^3$/s, 0.034, and 0.067 for the training period, and 10.1 m$^3$/s, 0.006, and 0.020 for the validation period. In addition, Figure 4 delineates one-to-one plots between the observed and simulated flow discharges by OrigLSTM, ParaLSTM with HIDL = 120, and ConcLSTM with HIDL = 120 for the training and validation periods. Specifically, all the points of one-to-one plots by ConcLSTM were very close to the one-to-one line for the training period.

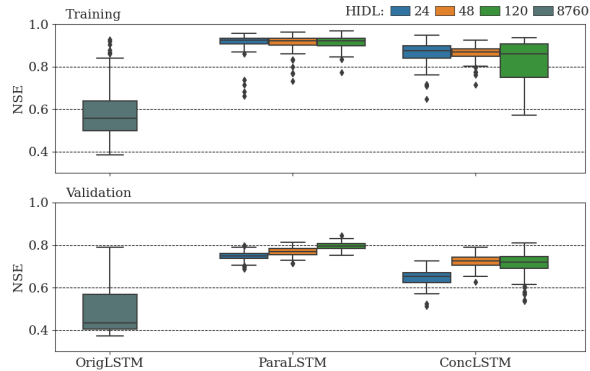

Figure 5. Nash-Sutcliffe efficiencies of the trained results by OrigLSTM, ParaLSTM, and ConcLSTM with different lengths of hourly time-series for the training period (top) and validation period (bottom).

To reiterate, the training process was conducted 100 times using each approach with each HIDL. For the training and validation periods, there were 100 trained results for each approach with each HIDL. For instance, the NSE value was calculated for each of the 100 trained results for both periods to investigate the stability of the training for each approach, as shown in Figure 5. OrigLSTM exhibits substantial differences in NSE values for the training and validation periods among the 100 training results. The difference in NSE between the best and median values were 0.373 and 0.356 for the training and validation periods, respectively. In contrast, differences in NSE values by ParaLSTM were minimal for each HIDL. Furthermore, the median values of NSE by ParaLSTM with all the HIDLs were comparable to the best value of NSE by OrigLSTM for the training and validation periods. The best value of NSE by OrigLSTM was 0.928 for the training period, whereas the median value of NSE was more than 0.920 by ParaLSTM. For the validation period, the best value of NSE by OrigLSTM was 0.789, whereas the median value by ParaLSTM with HIDL = 120 was 0.796. The results indicate that the ParaLSTM can be stably trained well. In addition, the training process of ConcLSTM with three HIDLs was more stable than that of OrigLSTM, although the NSE values by ConcLSTM were often smaller than those by ParaLSTM for the training and validation periods. For ConcLSTM, differences in NSE between the best and median values were the widest with HIDL = 120, which are 0.077 and 0.088 for the training and validation periods, respectively. ConcLSTM can yield high accuracy more stably than OrigLSTM.

## 5. Discussion

The results indicate that LSTM can learn long-duration dependencies between input and target time series variables even when IDL is substantial. However, LSTM with a considerable IDL requires substantial time for training. In addition, the training process is unstable. When the early stopping method was used, which stops the training process based on overfitting, the number of epochs until the early stopping method terminates the training process fluctuates. In addition, LSTM is not always adequately trained. Thus, LSTM experiences challenges with a considerable IDL.

The proposed approaches: ParaLSTM and ConcLSTM, significantly decrease the computational time for training the model compared to the original LSTM. Notably, Time/EP of LSTM significantly depends on IDL. The use of daily scale time-series data together with the hour-scale of those largely decreases IDL that is required to learn the long-duration dependencies between the input and the target time-series variables. For instance, when one year (365 days) of the input variable influences the target variable, ParaLSTM requires an IDL of 365 (days) whereas OrigLSTM requires 8760 (hours), that is, IDL becomes one 24th. Consequently, Time/EP is approximately 1/21.7. Because ConcLSTM has a longer IDL than ParaLSTM, Time/EP for ConcLSTM is longer than that for ParaLSTM. However, Time/EP for ConcLSTM remains at least 1/18 of that for OrigLSTM.

In addition, NEpochs is 27% - 49% smaller for ParaLSTM than for OrigLSTM on average. Because Time/EP is 1/21.7 for ParaLSTM than that for OrigLSTM, the total required computation time for the training becomes approximately 1/27.7-1/32.4 on average using ParaLSTM. ConcLSTM does not have an apparent advantage in terms of the number of NEpochs over OrigLSTM, except for HIDL=48. However, to reiterate, ConcLSTM requires much less computational time at each epoch than OrigLSTM. The total computational time for the training was still much less for ConcLSTM than for OrigLSTM. These results exhibit the large advantage of the required computation time when using both approaches: ParaLSTM and ConcLSTM.

ParaLSTM also yields advantages in terms of model accuracy and the stability of the training process over OrigLSTM in addition to the required computational resources. ParaLSTM with HIDL = 120 yielded higher model accuracy than OrigLSTM for the three periods (Table 1 and Figure 4). Meanwhile, most of the training results yielded high accuracy for the training and validation periods (Figure 5). For ParaLSTM, the finer (hourly) and coarser (daily) temporal resolutions of the input variables are provided parallel to the model, as shown in Figure 2a. There are some inconsistencies in the input data sequences over time. However, these inconsistencies do not worsen the model accuracy or the required computational time. Deep learning, including LSTM, is a black-box model. Unlike physical or conceptual models, it does not require consistency in the time increment between the input data. The results indicate that the flexible use of input data can improve the accuracy and required computational resources.

Conversely, ConcLSTM did not exhibit better evaluation metrics than OrigLSTM for the test period, although it did improve the model accuracy for the training and validation periods (Table 1). However, ConcLSTM does not have such inconsistencies over time, although the time increment of the input time-series changes from daily to hourly. Notably, to avoid such inconsistencies over time on ParaLSTM, ConcLSTM could be a reasonable choice because it still has an enormous advantage on the required computational time than OrigLSTM. In addition, the accuracy of ConcLSTM is still reasonable. The NSE value for ConcLSTM with HIDL = 120 is 0.748, which is more than 0.5, and is deemed acceptable, according to Moriasi et al. (2007).

## 6. Conclusions

This study proposed two approaches, namely ParaLSTM and ConcLSTM, to reduce the required computational time for the training of rainfall-runoff modeling in a snow-dominated watershed by RNN using multi-time-scale data as the input. Although ParaLSTM depicts inconsistencies over time among the input time-series, it significantly improves the required computational time for training. Furthermore, it improves the stability of the training and the accuracy of the simulated results. Although ConcLSTM also significantly improved the required computational time for training, it did not improve the accuracy. In contrast to ParaLSTM, ConcLSTM maintains consistency between the input time-series. Although ConcLSTM does not exhibit an advantage in terms of accuracy, it can still be useful owing to its reduced computational time compared to the original OrigLSTM approach. Finally, the proposed approaches can be extended to other RNN variants and other time scales (i.e., other than daily and hourly input and target series). In essence, the proposed approaches are straightforward with the potential of application in various time-series modeling problems.

### Acknowledgement

This work was supported by JSPS KAKENHI Grant Number 20K21916.

### References

Bowes, B.D., Sadler, J.M., Morsy, M.M., Behl, M., Goodall, J.L., 2019. Forecasting groundwater table in a flåood prone coastal city with long short-term memory and recurrent neural networks. Water 11, 1098. https://doi.org/10.3390/w11051098.

D. N. Moriasi, J. G. Arnold, M. W. Van Liew, R. L. Bingner, R. D. Harmel, T. L. Veith, 2007. Model evaluation guidelines for systematic quantification of accuracy in watershed simulations. Trans. ASABE 50, 885–900. https://doi.org/10.13031/2013.23153.

Gers, F.A., Schmidhuber, J., Cummins, F., 2000. Learning to forget: Continual prediction with LSTM. Neural Comput. 12, 2451–2471. https://doi.org/10.1162/089976600300015015.

Hochreiter, S., Schmidhuber, J., 1997. Long short-term memory. Neural Comput. 9, 1735–1780. https://doi.org/10.1162/neco.1997.9.8.1735.

Hrnjica, B., Bonacci, O., 2019. Lake level prediction using feed forward and recurrent neural networks. Water Resour. Manag. 33, 2471–2484. https://doi.org/10.1007/s11269-019-02255-2.

Hu, R., Fang, F., Pain, C.C., Navon, I.M., 2019. Rapid spatio-temporal flood prediction and uncertainty quantification using a deep learning method. J. Hydrol. 575, 911–920. https://doi.org/10.1016/j.jhydrol.2019.05.087.

Ishida, K., Tsujimoto, G., Ercan, A., Tu, T., Kiyama, M., Amagasaki, M., 2020. Hourly-scale coastal sea level modeling in a changing climate using long short-term memory neural network. Sci. Total Environ. 720, 137613. https://doi.org/10.1016/j.scitotenv.2020.137613.

Jeong, J., Park, E., 2019. Comparative applications of data-driven models representing water table fluctuations. J. Hydrol. 572, 261–273. https://doi.org/10.1016/j.jhydrol.2019.02.051.

Jeong, J., Park, E., Chen, H., Kim, K.Y., Shik Han, W., Suk, H., 2020. Estimation of groundwater level based on the robust training of recurrent neural networks using corrupted data. J. Hydrol. 582. https://doi.org/10.1016/j.jhydrol.2019.124512, http://www.ncbi.nlm.nih.gov/pubmed/124512.

Kao, I.-F., Zhou, Y., Chang, L.-C., Chang, F.-J., 2020. Exploring a Long Short-Term Memory based Encoder-Decoder Framework for Multi-Step-Ahead Flood Forecasting. J. Hydrol. 583. https://doi.org/10.1016/j.jhydrol.2020.124631, http://www.ncbi.nlm.nih.gov/pubmed/124631.

Kingma, D.P., Ba, J., 2014. Adam: A Method for Stochastic Optimization.

Kratzert, F., Klotz, D., Brenner, C., Schulz, K., Herrnegger, M., 2018. Rainfall-runoff modelling using Long Short-Term Memory (LSTM) networks. Hydrol. Earth Syst. Sci. 22, 6005–6022. https://doi.org/10.5194/hess-22-6005-2018.

Kratzert, F., Klotz, D., Herrnegger, M., Sampson, A.K., Hochreiter, S., Nearing, G.S., 2019. Toward improved predictions in ungauged basins: Exploiting the power of machine learning. Water Resour. Res. 55, 11344–11354. https://doi.org/10.1029/2019WR026065.

Li, W., Kiaghadi, A., Dawson, C.N., 2020. High temporal resolution rainfall runoff modelling using long-short-term-memory (LSTM). Networks.

Liang, C., Li, H., Lei, M., Du, aQ., 2018. Dongting Lake water level forecast and its relationship with the three Gorges Dam based on a long short-term memory network. Water 10, 1389. https://doi.org/10.3390/w10101389.

Liu, M., Huang, Y., Li, Z., Tong, B., Liu, Z., Sun, M., Jiang, F., Zhang, H., 2020. The applicability of LSTM-KNN model for real-time flood forecasting in different climate zones in China. Water 12, 440. https://doi.org/10.3390/w12020440.

Miao, Q., Pan, B., Wang, H., Hsu, K., Sorooshian, S., 2019. Improving monsoon precipitation prediction using combined convolutional and long short term memory neural network. Water 11, 977. https://doi.org/10.3390/w11050977.

Misra, S., Sarkar, S., Mitra, P., 2018. Statistical downscaling of precipitation using long short-term memory recurrent neural networks. Theor. Appl. Climatol. 134, 1179–1196. https://doi.org/10.1007/s00704-017-2307-2.

Paszke, A., Gross, S., Massa, F., Lerer, A., Bradbury, J., Chanan, G., Killeen, T., Lin, Z., Gimelshein, N., Antiga, L., Desmaison, A., Kopf, A., Yang, E., DeVito, Z., Raison, M., Tejani, A., Chilamkurthy, S., Steiner, B., Fang, L., Bai, J., Chintala, S., 2019. PyTorch: An imperative style, high-performance deep learning library, in: Wallach, H., Larochelle, H., Beygelzimer, A., d\textquotesingle Alché-Buc, F., Fox, E., Garnett, R. (Eds.). Adv. Neural Inf. Process. Syst. 32. Curran Associates, Inc., 8024–8035.

Song, T., Ding, W., Wu, J., Liu, H., Zhou, H., Chu, J., 2019. Flash flood forecasting based on long short-term memory networks. Water 12, 109. https://doi.org/10.3390/w12010109.

Tian, Y., Xu, Y.-P., Yang, Z., Wang, G., Zhu, Q., 2018. Integration of a parsimonious hydrological model with recurrent neural networks for improved streamflow forecasting Water 10, 1655. https://doi.org/10.3390/w10111655.

Tran Anh, D., Van, S.P., Dang, T.D., Hoang, L.P., 2019. Downscaling rainfall using deep learning long short-term memory and feedforward neural networks Int. J. Climatol. 39, 4170–4188. https://doi.org/10.1002/joc.6066.

Xiang, Z., Yan, J., Demir, I., 2020. Rainfall-runoffrunoff model with LSTM-based sequence-to-sequence learning Water Resour. Res. 56. https://doi.org/10.1029/2019WR025326.

Zhang, D., Lin, J., Peng, Q., Wang, D., Yang, T., Sorooshian, S., Liu, X., Zhuang, J. 2018. Modeling and simulating reservoir operation using an artificial neural network, support vector regression, and a deep learning algorithm.

Draft of Ishida, K., Kiyama, M., Ercan, A., Amagasaki, M., Tu, T., 2021.. J. hydroinformatics. https://doi.org/10.2166/hydro.2021.095

J. Hydrol. 565, 720–736. https://doi.org/10.1016/j.jhydrol.2018.08.050.

Zhang, J., Zhu, Y., Zhang, X., Ye, M., Yang, J., 2018. Developing a long short-term memory (LSTM)-based model for predicting water table depth in agricultural areas J. Hydrol. 561, 918–929. https://doi.org/10.1016/j.jhydrol.2018.04.065.

Zhu, S., Luo, X., Yuan, X., Xu, Z., 2020. Improved long short-term memory network for streamflow forecasting in the upper Yangtze River Stoch. Environ. Res. Risk Assess. 34, 1313–1329. https://doi.org/10.1007/s00477-020-01766-4.